\documentclass[aps,prb,showpacs,reprint]{revtex4-1}

\usepackage{graphicx}
\usepackage{amsmath}
\usepackage{bbm}

\begin{document}
\title
{Orbital effect on the in-plane critical field in free-standing superconducting nanofilms}
\author{P. W\'ojcik}
\email[Electronic address: ]{pawelwojcik@fis.agh.edu.pl}
\affiliation{AGH University of Science and Technology, Faculty of
Physics and Applied Computer Science, al. Mickiewicza 30,
Krak\'ow, Poland}
\author{M. Zegrodnik}
\affiliation
{AGH University of Science and Technology, Academic Centre for
Materials and Nanotechnology, al. A. Mickiewicza 30, Krak\'ow, Poland}

\begin{abstract}
The superconductor to normal metal phase transition induced by the in-plane magnetic field is
studied in free-standing Pb(111) nanofilms. In the considered structures the energy quantization
induced by the confinement leads to the thickness-dependent oscillations of the critical field
(the so-called 'shape resonances'). In this paper we examine the influence of the orbital effect on
the in-plane critical magnetic field in nanofilms. We demonstrate that the orbital term suppresses
the critical field and reduces the amplitude of
the thickness-dependent critical field oscillations. Moreover, due to the orbital effect,
the slope $H_{c,||}-T_c$ at $T_c(0)$ becomes finite and decreases with increasing film thickness in
agreement with recent experiments. The temperature $t^*$ at which the superconductor to normal metal
phase transition becomes of the first order is also analyzed.
\end{abstract}

\maketitle   

\section{Introduction}
The huge progress in nanotechnology which has been made in the last decade reopens the issue of
superconducting properties of
metallic nanostructures i.e.
nanofilms~\cite{Pfenningstorf2002,Zhang2010,Uchihashi2011,Qin2009,Wojcik2014_1},
nanowires~\cite{Tian2005,Zgirski2005,Shanenko2006} or metallic
grains~\cite{Bose2010,Garcia2008,Brun2009}. The studies of the
quantum size effect and its influence on the paired phase in thin films was initiated by Blatt and
Thomson in 1963~\cite{Blatt1963}. The main finding of their work~\cite{Blatt1963} was the appearance
of sharp oscillations of the critical temperature as a function of the film thickness. As it was
argued~\cite{Blatt1963} this effect results from the confinement of the electron motion in the
direction perpendicular to the film. If the size of the system becomes comparable to the electron
wave length, the Fermi sphere splits into a set of discrete two-dimensional subbands, energy of
which increases with decreasing film thickness. Each time a bottom of a subband passes through the
Fermi level, a sharp peak of the critical temperature appears. Due to technological difficulties in
the preparation of uniform films, which were typically polycrystalline and contained a large number
of defects, the experimental observation of the so-called shape resonances has been reported only
recently~\cite{Ozer2006,Ozer2007,Guo2004,Eom2006}. Measurements of the critical temperature for
Pb(111) nanofilms grown on Si substrate~\cite{Guo2004,Eom2006} revealed that the period of the
thickness-dependent oscillations is equal to $\sim2$~ML. Additionally, the beating effect with the
periodicity varying from $7$~ML to $14$~ML was observed. This feature, called bilayer or even-odd
oscillations, was theoretically studied by Shanenko et al. in Ref.~\cite{Shanenko2007}. 
The studies of superconducting properties of the ultra-thin films have been recently extended to the
case of high temperature and multiband superconductors. Recently, the enhancement of the
superconducting critical temperature with respect to the bulk limit has been reported for interfaces
and heterostuctures based on cuprates~\cite{Gozar2008}, iron pnictides ~\cite{Liu2012} and
LaAlO$_3$/SrTiO$_3$~\cite{Reyren2007}. Moreover, the experimental reports on the growth techniques
of high quality MgB$_2$ films of thicknesses less than 10~$nm$ ~\cite{Zhang2011,Zhang2013} entailed
a series of theoretical papers describing the quantum size effect in multiband thin film
superconductors ~\cite{Szawlowski2006,Innocenti2010,Innocenti2011,Araujo2011,Romero2014}.

Recent studies devoted to superconductivity in the nanoscale regime concern the effect of
the quantum confinement on the superconductor to normal metal phase transition induced by the
magnetic field~\cite{Shanenko2008}. The thickness-dependent oscillations of the perpendicular and
parallel critical field for ultra-thin lead films were reported by Bao et al. in
Ref.~\cite{Bao2005}. Moreover, the study of the superconductor to normal metal transition 
induced by the parallel magnetic field for Pb monolayer was recently presented by Sekihara et al. in
Ref.~\cite{Sekihara2013}. 
In both experiments, the measured parallel critical field was higher than the Pauli paramagnetic
limit. This unusual behavior  has been explained in our recent 
papers~\cite{Wojcik2014_3,Wojcik2014_2} in which we have investigated the quantum size effect on the
in-plane critical field in paramagnetic limit. We have shown~\cite{Wojcik2014_3} that the
zero-temperature critical field for nanofilms is higher than  the Clogston - Chandrasekhar (CC)
paramagnetic limit and diverges to the CC (Pauli) limit for sufficiently thick films. This fact has
been explained on the basis of the spatially varying energy gap induced by the confinement. In
Ref.~\cite{Wojcik2014_3} the new formula for the paramagnetic critical field in nanofilms  has been
proposed. However, the analysis presented in
Ref.~\cite{Wojcik2014_3} has been carried out in the paramagnetic limit. The extension of this study
for films thicker than $15$~ML requires the inclusion of the orbital effect which significantly
affects the superconductor to normal metal transition. According to our knowledge such study has not
been reported until now. 

In the present paper we consider free-standing Pb(111) metallic nanofilms and investigate the
orbital effect on the superconductor-normal metal transition driven by the in-plane magnetic field.
Based on the analysis of the spatially  dependent energy gap we study the
influence of the orbital effect on the critical magnetic field oscillations induced by the
confinement. The analysis of the thermal effect in terms of the orbital effect is also included.
The paper is organized as follows: in Sec.~\ref{sec:model} we introduce the basic concepts of the
theoretical scheme based on the BCS theory, in Sec.~\ref{sec:results} we present the results
while the summary is included in Sec.~\ref{sec:concl}.

\section{Theoretical method}
\label{sec:model}
The phonon-mediated superconductivity in metallic nanofilms can be described with the use of the BCS
theory. The Hamiltonian of the system is given by
\begin{eqnarray}
\hat{\mathcal{H}}&=&\sum _{\sigma} \int d^3 r \:
\hat{\Psi}^{\dagger} (\mathbf{r},\sigma) \hat{H}_e ^{\sigma}
\hat{\Psi}(\mathbf{r},\sigma) \nonumber \\ 
&+& \int d^3 r \left [ \Delta
(\mathbf{r})\hat{\Psi}^{\dagger}(\mathbf{r},\uparrow)
\hat{\Psi}(\mathbf{r},\downarrow) +H.c. \right ] \nonumber \\
&+&\int d^3r \frac{|\Delta(\mathbf{r})|^2}{g},
\label{eq:ham}
\end{eqnarray}
where $\sigma$ indexes the spin state $(\uparrow, \downarrow)$ and $g$ is the electron-phonon
coupling. In the presence of the in-plane magnetic field $H_{||}$, the single-electron Hamiltonian
$\hat{H}_e^{\sigma}$ can be expressed as
\begin{equation}
\hat{H}_e ^\sigma = \frac{1}{2m} \left ( -i\hbar \nabla +\frac{e}{c}
\mathbf{A} \right )^2 + s\mu_B H_{||} - \mu _F,
\label{eq:hame}
\end{equation}
where $s=+1(-1)$ for $\sigma=\uparrow (\downarrow)$, $m$ is the effective electron mass, $\mu_F$ is
the chemical potential, $\mathbf{A}=(0,-H_{||}z,0)$ is the vector potential corresponding to
the magnetic field $\mathbf{H}=(H_{||},0,0)$ applied in-plane and the energy gap
$\Delta(\mathbf{r})$ is
defined as
\begin{equation}
 \Delta(\mathbf{r})=-g \left < \hat{\Psi} (\mathbf{r},\downarrow)
\hat{\Psi}^{\dagger} (\mathbf{r},\uparrow)  \right >.
\end{equation}
In ultrathin nanofilms the electron motion is limited in the direction perpendicular to the film
($z$ axis) resulting in the quantization of the electron energy. We assume that the system is
infinite in the $x-y$ plane. Thus, the field operators in
Eq.(\ref{eq:ham}) are expressed as
\begin{eqnarray}
 \hat{\Psi}(\mathbf{r},\sigma)=\sum_{n,\mathbf{k}}
\phi_{\mathbf{k}n}(\mathbf{r})  \:
\hat{c}_{\mathbf{k} n \sigma}, \\
\hat{\Psi}^{\dagger}(\mathbf{r},\sigma)=\sum_{\mathbf{k},n}
\phi^*_{\mathbf{k} n}(\mathbf{r})  \:
\hat{c}^{\dagger}_{\mathbf{k} n \sigma},
\end{eqnarray}
where $\hat{c}_{\mathbf{k} n \sigma}
(\hat{c}^{\dagger}_{\mathbf{k} n \sigma})$ is
the anihilation (creation) operator for an electron with spin $\sigma$ in a state characterized by
the quantum numbers $(\mathbf{k},n)$ while $\phi_{\mathbf{k} n}(\mathbf{r})$ is the single-electron
eigenfunction of the Hamiltonian $\hat{H}^{\sigma}_e$ whose
explicit form is given by
\begin{equation}
 \phi_{\mathbf{k}n}(\mathbf{r})=\frac{1}{2\pi} e^{ik_xx}e^{ik_yy} \varphi_{k_yn}(z).
\label{eq:ses}
\end{equation}
where $\mathbf{k}=(k_x,k_y)$ is the electron wave vector and $n$ labels the discrete quantum
states induced by the confinement along the $z$ axis. \\
By using the eigenfunctions given by Eq.~(\ref{eq:ses}), one can reduce the Hamiltonian (\ref{eq:hame}) to
the 1D form which corresponds to the $z$ dependent part - $\varphi_{k_yn}(z)$.
\begin{eqnarray}
 \hat{H}_{e,1D} ^\sigma &=& -\frac{\hbar ^2}{2m}\frac{\partial ^2}{\partial z^2} +\frac{\hbar
^2}{2m} \left ( k_x^2+k_y^2 \right ) - \frac{\hbar e H_{||}k_y}{m}z \nonumber \\ 
&+& \frac{e^2H_{||}^2}{2m}z^2 + s\mu_B H_{||} - \mu _F,
\label{eq:hame1D}
\end{eqnarray}
Note, that in the presence of the magnetic field $H_{||}$ the eigenfunctions $\varphi_{k_y n}(z)$
depend on the $k_y$ component of the wave vector. In our calculations  $\varphi_{k_yn}(z)$ are
determined numerically by the diagonalization of the Hamiltonian (\ref{eq:hame1D}) in the
basis of the quantum well states 
\begin{equation}
 \varphi_{k_yn}(z)= \sqrt{\frac{2}{d}} \sum _{l} c_{k_ynl} \sin \left [
\frac{\pi(l+1)z}{d} \right ],
\label{basis}
\end{equation}
where $d$ is the film thickness whereby we adopt the hard-wall confinement as
the boundary condition in the $z$ direction. \\
By using the Bogoliubov -- de Gennes transformation $\hat{c}_{\mathbf{k} n \sigma} =
u_{kn\sigma}\gamma_{kn}+s v^*_{kn\sigma}\gamma^{\dagger}_{kn}$~\cite{Gennes}, the energy gap in the
band $n$ defined as $\Delta_{n}=\langle \phi_{\mathbf{k}n} | \Delta(\mathbf{r}) |
\phi_{\mathbf{k}n} \rangle $ can
be expressed as follows
\begin{eqnarray}
&& \Delta_{n'}=\frac{g}{4 \pi ^2} \int d k_x d k_y \times \\
& \times & \sum _{n'} C_{k_yn'n}
\frac{\Delta_{n}}{2 \sqrt{\xi_{\mathbf{k}n}^2+\Delta_{n}^2}} \left [
1-f(E^{+}_{\mathbf{k}n}) - f(E^{-}_{\mathbf{k}n}) \right ] \nonumber
\label{deltan},
\end{eqnarray}
where $\xi_{\mathbf{k}n}$ is the single-electron energy, $E^{\pm}_{\mathbf{k}n}=\pm \sqrt{\xi
_{\mathbf{k}n} ^2 + |\Delta _n| ^2}$ is the quasi-particle energy, $f(E)$ is the Fermi-Dirac
distribution and $C_{k_y n'n}$ are the interaction-matrix elements
given by 
\begin{equation}
 C_{k_y n'n}=\int dz \varphi_{k_y n'}(z) \varphi_{-k_y n'}(z)
\varphi_{k_y n}(z) \varphi_{-k_y n}(z).
\label{eq_inter}
\end{equation}
The summation in Eq.~(\ref{deltan}) is carried out only
over the single-electron states with energy $\xi _{\mathbf{k} n}$
inside the Debye window $\left | \xi _{\mathbf{k} n}  \right | <
\hbar \omega _D$, where $\omega _D$ is the Debye frequency~\cite{Gennes}.\\
The spatially dependent order parameter $\Delta(z)$ can be
expressed as
\begin{eqnarray}
\Delta(z)&=&\frac{g}{4 \pi^2} \int d k_x d k_y 
\sum_{n} \varphi_{k_y n}(z) \varphi_{-k_y n}(z) \times \nonumber \\
&\times&
\frac{\Delta_{n}}{2 \sqrt{\xi_{\mathbf{k} n}^2+\Delta_{n}^2}}
\left [ 1-f(E^{+}_{\mathbf{k} n})-f(E^{-}_{\mathbf{k} n}) \right ].
\label{delta}
\end{eqnarray}
The energy gap $\Delta(z)$ for given nanofilm thickness $d$ is calculated in the self-consistent
manner using the following procedure. For each value of the wave vector $k_y$ (note that the range
of the wave vector is limited by the condition  $\left | \xi _{\mathbf{k} n}  \right | <
\hbar \omega _D$) we calculate the single-electron wave functions $\varphi_{k_yn}(z)$ and
the energies $\xi_{\mathbf{k}n}$ by the diagonalization of the Hamiltonian (\ref{eq:hame}) in the
basis given by Eq.~(\ref{basis}). Then, the single-electron wave functions $\varphi_{k_yn}(z)$ are
used to determine the staring value of $\delta _n$. The final value of $\delta _n$ is calculated
by the self-consistent procedure given by Eq.~\ref{eq_inter}. Finally, having $\delta _n$ and
$\varphi_{k_yn}(z)$, the spatially dependent order parameter $\Delta(z)$ is calculated by the use of
Eq.~\ref{delta}. The calculations in the paramagnetic limit (with no orbital effect) are carried out
by putting $H_{||}=0$ in Eq.~\ref{eq:hame1D} everywhere except the Zeeman term. In this case the
sum (\ref{basis}) reduces to a single term - the situation corresponds to a infinite quantum well
with the spin Zeeman splitting.

We should note that aforementioned procedure can lead to solutions with $\Delta \neq 0$ even for the
values of the magnetic field for which the free energy of the superconducting state is greater than
the free energy corresponding to the normal metal solution ($\Delta=0$). Therefore, for each value
of the magnetic field we calculate and compare the free energies of the normal and superconducting
phase. The theoretical  study of the free energy in superconducting nanostructures has been
presented in detail in Ref.~\cite{Kosztin}.\\

Due to the confinement the chemical potential can strongly deviate from the bulk value. For this
reason for each nanofilm thickness we determine the chemical potential from the formula
\begin{eqnarray}
 n_e&=&\frac{1}{d} \int dk_x dk_y \sum _{n \sigma}\int _0^d dz
\bigg \{ |u_{\mathbf{k}n\sigma}\varphi_{k_yn}(z) |^2f(E_{\mathbf{k}n})  \nonumber \\ 
 &+&|v_{\mathbf{k}n\sigma} \varphi_{k_yn}(z)|^2 [1-f(E_{\mathbf{k}n})]\bigg \},
\end{eqnarray}
where $n_e$ is the electron density corresponding to the bulk value (corresponding to the chemical
potential $\mu_{bulk}$).

\section{Results}
\label{sec:results}
In the present paper we consider the free-standing
Pb(111) nanofilms. The first-principle calculations of the
quantized band structure for Pb nanofilms in (111) and (100) directions
are presented in Refs.~\cite{Wei2002,Wei2007,Miller2009}. Authors
of these papers have pointed out that in (111) direction the energy
dispersion is nearly parabolic and the quantum size effect can be well
described by the quantum well states centered at the L-point of a
two-dimensional Brillouin zone~\cite{Wei2002}. 
Based on these results, in our analysis we use
the parabolic band approximation treating the bulk Fermi level $\mu
_{bulk}$ and the electron mass $m$ as the fitting parameters. Their
values are determined based on the results from the first-principle
calculations for Pb(111) presented in
Refs.~\cite{Wojcik2014_3,Wei2002}. 
We use the following values of the parameters:
$gN_{bulk}(0)=0.39$ where $N_{bulk}(0)=mk_F/(2 \pi^2 \hbar ^2)$  is the
bulk density of the single-electron states at the Fermi level, $\hbar
\omega _D=8.27$~meV, the bulk energy gap
$\Delta_{bulk}=1.3$~meV and $\mu _{bulk}=3.8$~eV which
corresponds to the electron density $n_e=4.2 \times 10^{21}$~cm$^{-3}$.\\

\subsection{Orbital effect on the critical magnetic field.}
\label{ra}
\begin{figure}[ht]
\begin{center}
\includegraphics[scale=0.38]{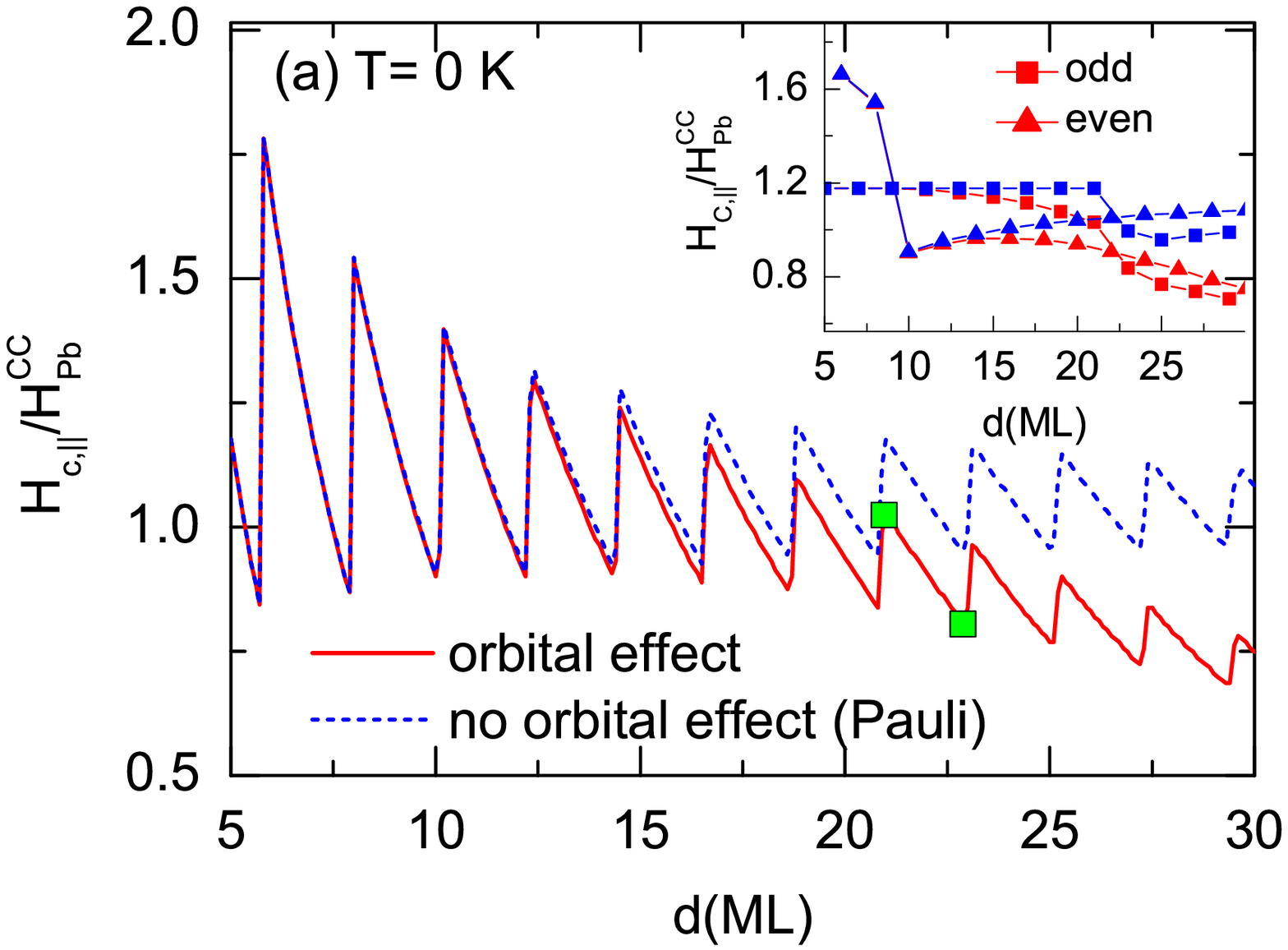}
\includegraphics[scale=0.38]{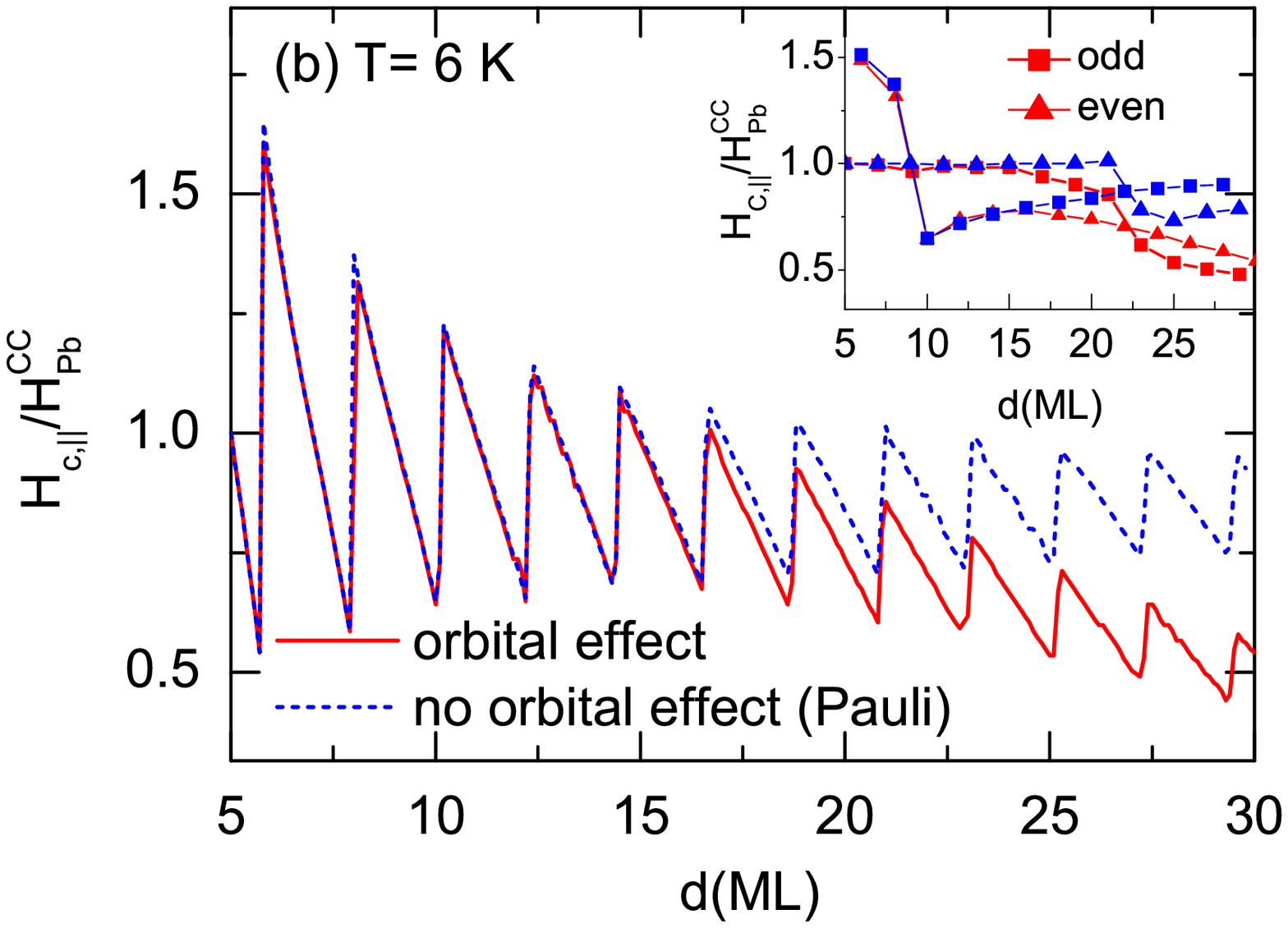}
\caption{(Color online) In-plane critical magnetic field $H_{c,||}$ in units of $H^{CC}_{Pb}$ as a
function of the nanofilm thickness $d$ calculated with and without
(Pauli approximation) the inclusion of the orbital effect. Results
for the temperature (a)~$T=0$~K and (b)~$T=6$~K. 
Insets present $H_{c,||}(d)$ as a functions of number of Pb
monolayers assuming the lattice constant $a_{Pb}=0.286$~nm.
Green squares in panel (a) denote the film thicknesses chosen for presentation
in Fig.~\ref{fig5}.
}
\label{fig1}
\end{center}
\end{figure}

Figure ~\ref{fig1} displays the in-plane critical field $H_{c,||}$ in units of $H^{CC}_{Pb}$
as a function of the nanofilm thickness $d$, where $H^{CC}_{Pb}$ is the paramagnetic critical field
calculated on the basis of the Clogston-Chandrasekhar limit~\cite{Clogston1962,Chandrasekhar1962}
($H^{CC}=\Delta_{bulk}/\sqrt{2} \mu _B$) which for bulk Pb gives $H^{CC}_{Pb}=15.9$~T. 
The thickness range under consideration is chosen on the basis of the experiments which
report the stable Pb(111) nanofilms with the thickness varying from 5 ML
to 30 ML~\cite{Guo2004,Eom2006}. Within our analysis the value of $H_{c,||}$ for each nanofilm
thickness is defined as the field for which the spatially averaged energy gap
$\bar{\Delta}=(1/d)\int_0^d\Delta(z)dz$ drops below $0.01\Delta _{bulk}$.
The physical origin of 'the tooth-like' oscillations  in Fig.~\ref{fig1}
can be explained in terms of the electron energy quantization related to the confinement of
the electron motion in the direction perpendicular to the film~\cite{Wojcik2014_1,Shanenko2008}.
If the thickness of the nanofilm becomes comparable with the electrons wave length the Fermi sphare
splits into the set of discrete subbands. Increase of the film thickness results in a decrease of
the discrete subband energies. Each time a subband passes through the Fermi level
the density of states in the energy window $\left [ \mu - \hbar \omega
_D,  \mu + \hbar \omega _D \right ]$, in which the phonon-mediated pairing occurs,
abruptly increases. This phenomenon leads to the thickness-dependent oscillations of the energy gap
and the critical field presented in Fig.~\ref{fig1}. As we can see, in the thickness range between
two subsequent resonances, the critical magnetic field almost linearly decreases with increasing
nanofilm thickness~\cite{Blatt1963,Shanenko2008}.
The predicted enhancement of the critical field reaches almost twice the value of the 
paramagnetic limit in the bulk Pb. Similar behavior was recently reported for Pb nanofilms in
Ref.~\cite{Sekihara2013} where it is shown that experimentally measured critical field $H_{C,||}$ was
much higher than the paramagnetic limit $H^{CC}_{Pb}$. The increase of the paramagnetic
in-plane critical field in nanofilms has been explained in our recent paper~\cite{Wojcik2014_3}.
In the insets of Fig.~\ref{fig1} we show the value of $H_{c,||}$ as a function of	
monolayer number (we assume the lattice constant $a_{Pb}=0.286$~nm corresponding to the bulk value).
Results presented in such manner reveal the bilayer (even-odd) oscillations with the beating
effect observed in experiments with Pb nanofilms~\cite{Guo2004,Eom2006} and explained in
Refs.~\cite{Shanenko2007,Wojcik2014_3}.

\begin{figure}[ht]
\begin{center}
\includegraphics[scale=0.4]{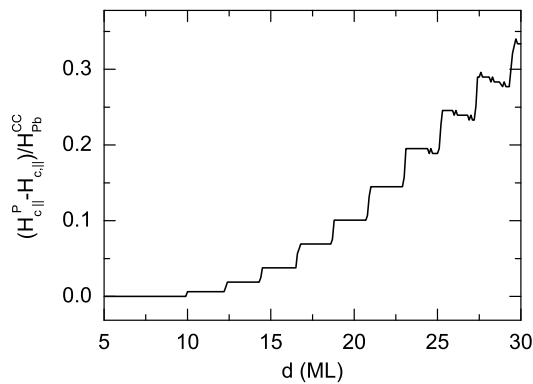}
\caption{(Color online) Difference between the critical magnetic field
calculated in the Pauli approximation ($H^P_{c,||}$) and with the inclusion of the orbital effect
($H_{c,||}$) in units of $H^{CC}_{Pb}$.
Results for $T=0$~K.
}
\label{fig2}
\end{center}
\end{figure}

In order to analyze the influence of the orbital effect on the in-plane critical field, in
Fig.~\ref{fig1} we present the results of the calculations carried out with the inclusion of the
orbital effect (solid, red line) and in the Pauli approximation (dashed, blue line).
From Fig.~\ref{fig1} one can see that the orbital
effect leads to a decrease of the critical field what is clearly visible for thick nanofilms.
In comparison with the Pauli approximation the amplitude of the critical field
oscillations is also reduced with increasing nanofilm thickness. One should note (see
Fig.\ref{fig2}) that the orbital effect significantly affects the value of $H_{c,||}(d)$ for
nanofilms with the thickness greater than $15$~ML.
The upper limit of the  thickness above which the Pauli approximation is
no longer satisfied can be approximated by the magnetic
length $a_H=\sqrt{\hbar / e H^{P}_{||}}$, where $H^{P}_{||}$ is the
paramagnetic (Pauli) critical field. 
Nevertheless, the use of the Clogston-Chandrasekhar paramagnetic field for the bulk
$H^{CC}_{Pb}=15.9$~T gives $a_H\approx22$~ML which is greater that the thickness limit $15$~ML
determined from the numerical calculations. This discrepancy results from the different values of
the paramagnetic critical field $H_{||}^P$ in nanofilms in comparison with the bulk
value~\cite{Wojcik2014_3}.

The suppression  of the critical field induced by the orbital effect can
be explained on the basis of the classical Lorentz force acting on
electrons in the magnetic field.
\begin{figure}[ht]
\begin{center}
\includegraphics[scale=0.38]{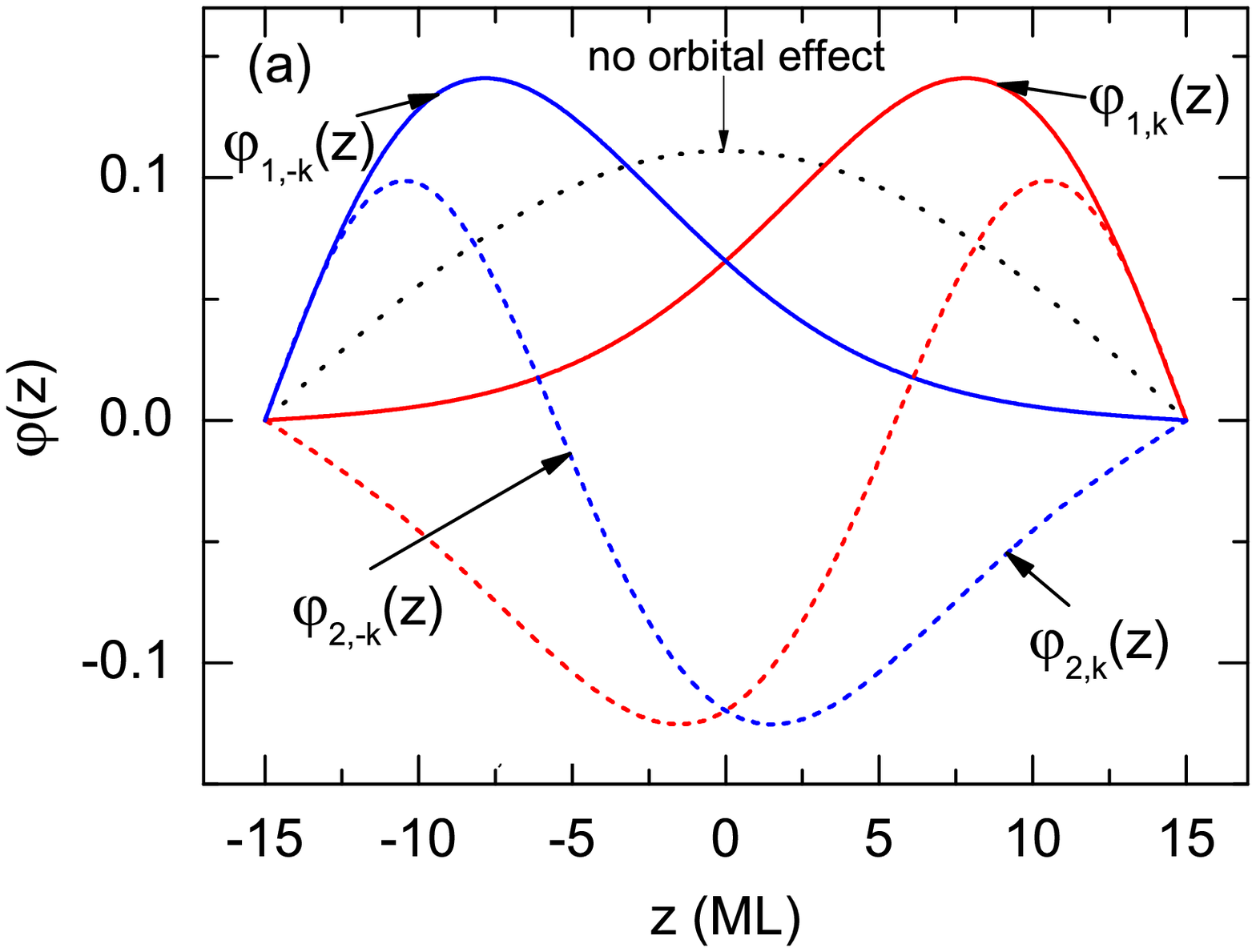}
\includegraphics[scale=0.38]{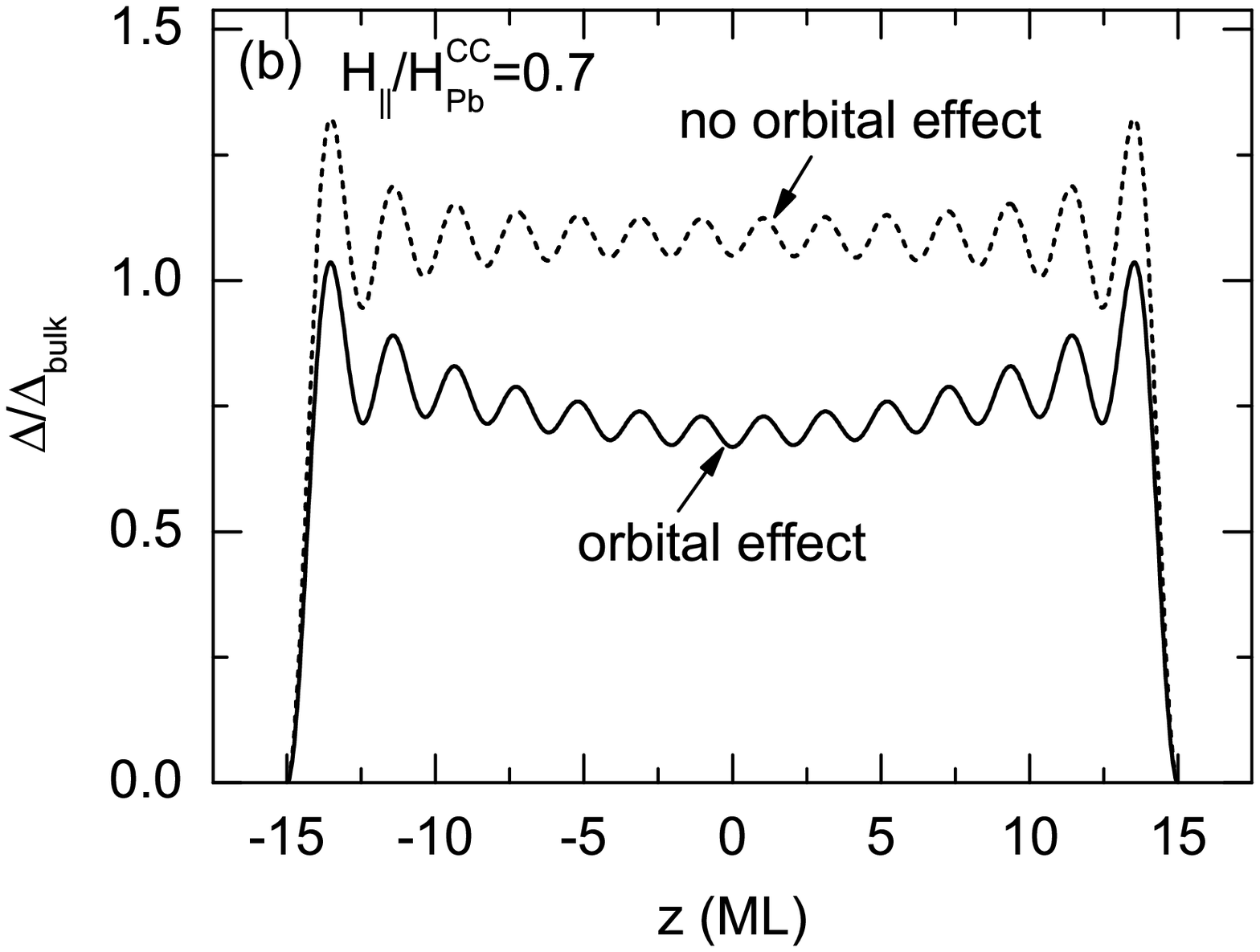}
\caption{(Color online) (a) Wave  functions for the first 
$\varphi_{1,\pm k}(z)$  and the second  $\varphi_{2,\pm k}(z)$
quantum well states calculated for opposite $\mathbf{k}$ vectors. States
with opposite $\mathbf{k}$ are shifted towards the opposite edges of the
sample. The wave function $\varphi_{1,+k}(z)=\varphi_{1,-k}(z)$
obtained for the case with no orbital effect 
is marked by the black, dotted line. (b)
Spatial dependent energy gap $\Delta(z)$ in units of $\Delta _{bulk}$ calculated for
the magnetic field $H_{||}/H^{CC}_{Pb}=0.7$ with (solid line) and without  (dashed
line) orbital effect. Results for $d=30$~ML.
}
\label{fig3}
\end{center}
\end{figure}
The parallel magnetic field  applied along the $x$ axis (for the
$s$-wave superconductors the direction of $H_{||}$ is not relevant)
results in the Lorentz force directed perpendicular to the plane (along
the $z$ axis). 
For electrons with opposite $\mathbf{k}$ vector (which form the Cooper pair) the Lorentz forces have
opposite orientations.
As a consequence, for nanofilms with the thickness
reduced to several monolayers, the electron density in the $z$-direction
is shifted from the center to the edges (in accordance with the
Lorentz force orientation). 
The shift of the density for electrons with opposite $\mathbf{k}$
towards opposite edges of the sample is clearly visible in
Fig.~\ref{fig3}(a) which presents the single-electron wave functions for the first
$\varphi_{1,\pm k}(z)$ and the second $\varphi_{2,\pm k}(z)$ quantum
well state. 
This effect does not take place for the case with no orbital term included
for which $\varphi_{1,+k}(z)=\varphi_{1,-k}(z)$.
According to Eq.~(\ref{eq_inter}) the energy gap depends on the
overlap between the wave functions of electrons with opposite
momenta, which from the Cooper pairs. As presented in
Fig.~\ref{fig3}(a) the orbital effect results in the reduction of this
overlap and consequently leads to the suppression of the energy gap and
the critical magnetic field. We should mention that the presented explanation is correct only
for nanofilms which are too thin for the formation of the vortex
state, which is the case considered here. The Lorentz force 
manifest itself also in  the spatial
distribution of  the energy gap presented in Fig.~\ref{fig3}(b).  
It is well known that in superconducting nanostructures the energy
gap is not uniform, as in the bulk, but it depends on the position. The
spatially dependent energy gap for the nanofilm thickness $d=30$~ML and
the magnetic field $H_{||}/H^{CC}_{Pb}=0.7$ is presented in Fig.~\ref{fig3}(b).
The comparison of the results calculated with and without the inclusion of the orbital effect
allows to conclude that the reduction of the gap parameter caused by the orbital effect is most
significant at the center of the film. 
\begin{figure}[ht]
\begin{center}
\includegraphics[scale=0.4]{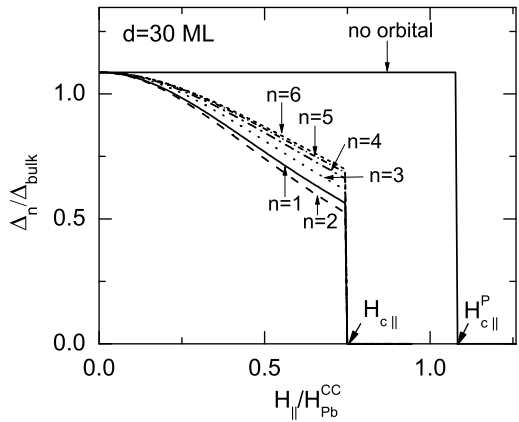}
\caption{Energy gap $\Delta_n$ in units of $\Delta_{bulk}$ as a function of the magnetic field
$H_{||}$ in units of $H^{CC}_{Pb}$ for the nanofilm thickness $d=30$~ML.
}
\label{fig4}
\end{center}
\end{figure}

Let us now explain the 'step-like' dependence of the difference
$H^P_{c,||}-H_{c,||}$ as a function of the film thickness presented in Fig.~\ref{fig2}. The
thicknesses $d$ for which the steps appear correspond to  peaks of the
critical field  seen in Fig.~\ref{fig1}. This indicates that the presented behavior is
directly related to the discrete energy spectrum induced by the confinement.
Therefore a subband passing through the Fermi level (with increasing $d$) not only contributes to
the increase of the density of state at the Fermi level but also to the orbital
effect. This orbital contribution from subsequent subbands manifests as the steps in
Fig.~\ref{fig2}. Since the orbital effect corresponding to each subband is different, the heights of
the steps also differ. 
The different importance of the orbital term in different subbands is clearly visible in
Fig.~\ref{fig4} in which we present the subband energy gap $\Delta_n$
as a function of the magnetic field $H_{||}$. 
For comparison $\Delta _n(H_{||})$ in the Pauli limit is also shown. The slow decrease of the energy
gap with increasing magnetic field seen in Fig.~\ref{fig4} results from the orbital effect. Note
that in the Pauli limit $\Delta_n$ is the same
for all subbands and does not depend on the magnetic field in the superconducting state.
This clearly show that the orbital effect coming from different subbands $n$ is
different resulting in the steps with unequal height in Fig.~\ref{fig2}.\\

\subsubsection{Orbital effect on $H_{c,||}(T_c)$}
\label{rb}
\begin{figure}[ht]
\begin{center}
\includegraphics[scale=0.5]{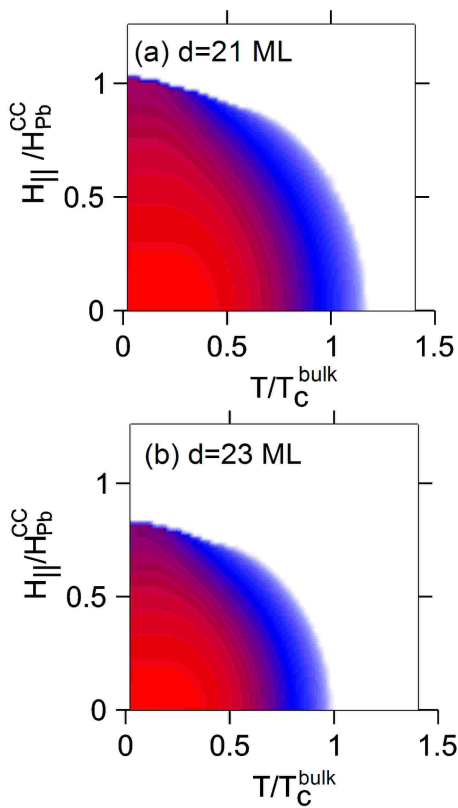}
\caption{(Color online) Spatially averaged energy gap $\bar{\Delta}$ as
a function of magnetic field $H_{||}$ and temperature $T$
for the nanofilm thickness (a) $d=21$~ML and (b) $d=23$~ML. The value of
the energy gap in each figure is 
normalized with respect to its maximum.
}
\label{fig5}
\end{center}
\end{figure}

\begin{figure}[ht]
\begin{center}
\includegraphics[scale=0.4]{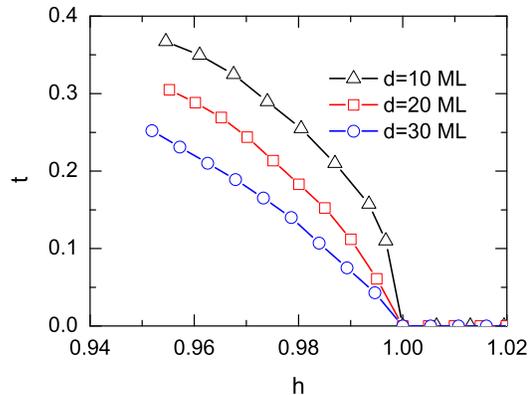}
\caption{(Color online) Normalized critical magnetic field $h=H_{c,||}/H_{c,||}(0)$ as a
function of normalized temperature $t=T/T_c(0)$ for different nanofilm thicknesses $d$.
}
\label{fig6}
\end{center}
\end{figure}

In this subsection we discuss the influence of the orbital effect 
on the superconductor to normal metal phase transition induced by the
magnetic field in non-zero temperature. 
In Fig.~\ref{fig5} we present the spatially averaged energy gap as a
function of magnetic field and temperature for the 
nanofilm thickness (a)~$d=21$~ML and (b)~$23$~ML. The chosen
thicknesses correspond to the maximum and the minimum of the
zero-temperature critical field presented in Fig.~\ref{fig1}(a).
As one can see the range of magnetic field and temperature in which
the film remains in the superconducting state varies with 
its thickness. Since the significant impact of the orbital effect can be
observed in the vicinity of $T=T_c(0)$, we restrict our analysis to
this range.
Fig.~\ref{fig6} presents the $h-t$ phase diagram for different nanofilm
thicknesses, where $h$ and $t$ are the
normalized critical magnetic field $h=H_{c,||}/H_{c,||}(0)$ and the
normalized temperature $t=T/T_c(0)$, respectively. $H_{c,||}(0)$ and
$T_c(0)$ are the critical field at $T=0$ and the critical temperature
for $H_{||}=0$.
As presented in Fig.~\ref{fig6}, for the nanofilm thickness $d=10$~ML
for which the orbital effect is negligibly small, the slope of $h(t)$ at
$t=1$ is infinite  
and can be approximated by the formula
$dh/dt \approx \sqrt{1/(1-t)}$~\cite{Ketterson}. Due to the orbital
effect the slope $h(t)$ at $t=1$ becomes finite for thicker
nanofilms ($d=20,30$~ML) and gradually decreases with increasing
thickness. Similar behavior has been recently reported
in the experiments with Pb nanofilms~\cite{Ozer2006,Ozer2006_b}. 
The authors of Refs.~\cite{Ozer2006,Ozer2006_b} have
argued that such dependence is caused by the boundary scattering, and therefore 
results from the roughness of the sample. As we have shown here
the same behavior can be also induced by the orbital effect in the ultra clean film.

\begin{figure}[ht]
\begin{center}
\includegraphics[scale=0.4]{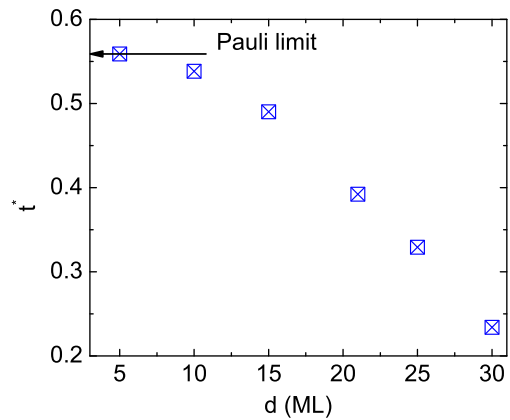}
\caption{(Color online) Temperature $t^*$ at which the superconductor
to normal metal transition becomes of the first order as a function of
the nanofilm thickness $d$.
}
\label{fig7}
\end{center}
\end{figure}

The orbital effect in nanofilms influences also the order of the
superconductor to normal metal phase transition. It is well known that
for the orbital limiting case, with no Pauli effect, the superconductor
to normal metal transition induced by the magnetic field is of the second
order. In contrast, in the Pauli limit the
second order transition is suppressed with decreasing temperature and
below $t^*=0.56$ it becomes of the first order~\cite{Maki1964,Maki1966}.
The superconducting nanofilms are systems in
which the orbital and the paramagnetic effects are comparable while their
relative importance can be controlled by changing the film thickness.
By analyzing the free energy of the superconducting and the normal state we have
determined the temperature $t^*$ at which the superconductor to
normal metal phase transition becomes of the first order for different
nanofilm thickness. In Fig.~\ref{fig7} one can see that for ultrathin
nanofilms, for which the paramagnetic effect is dominant, $t^*$
diverges to the value $0.56$ predicted by Kazumi Maki in
Refs.~\cite{Maki1964,Maki1966}. 
If we increase the thickness, the temperature $t^*$ decreases which is
directly related to the enhancement of the orbital effect. The
extrapolation of $t^*(d)$ for $t^*=0$ gives $d=39$~ML which is the limit
above which the transition of the second order occurs in the whole range
of temperatures which corresponds to the orbital limiting case.

\section{Summary}
\label{sec:concl}
The superconductor to normal metal phase transition driven by the
in-plane magnetic field for Pb(111) nanofilms has been investigated in
the framework of the BCS theory. We have shown that the orbital effect
suppresses the critical field as well as the amplitude of the critical
field oscillations induced by the quantum confinement
(shape-resonances). The analysis of the $H_{c,||}-T_c$
diagram allows to demonstrate that due to the orbital effect the slope
$H_{c,||}-T_c$ at $T_c(0)$ becomes finite and systematically decreases
with increasing film thickness. This result agrees with recent
experiments for Pb nanofilms~\cite{Ozer2006,Ozer2006_b}. 
We have also analyzed the thermal effect and shown that the
temperature $t^*$  at which the superconductor to normal metal
transition becomes of the first order reduces with increasing film
thickness. For the ultrathin nanofilms, in which the Pauli pair-breaking
mechanism is dominant, $t^*$ approaches to $0.56$ in agreement with the
theoretical prediction by Kazumi Maki~\cite{Maki1964,Maki1966}.

It is worth mentioning that our study take into account only the electronic structure. 
However, in the nanoscale regime, the confinement 
affects not only the electronic spectrum but also the phononic degrees of freedom which for
nanofilms strongly deviates from that in the bulk~\cite{Nabity1992}. The quantization of the
phononic spectra in nanofilms and its influence on superconducting properties have been considered
in many papers~\cite{Hwang2000,Zhang2005,Chen2013}. Moreover, the effect of the confinement on the
electron-phonon coupling strength has been recently studied  by Saniz et al. in
Ref.~\cite{Saniz2013}. In this paper~\cite{Saniz2013} attractive electron-electron interaction has
been derived with the use of the Green function approach beyond the contact potential approximation.
It has been found that the increase of the critical temperature observed in superconducting
nanofilms is due to the increase of the number of phonon modes what results in the enhancement of
the electron-phonon coupling. 
Since the modification of the phononic dispersion due to the confinement only slightly changes the
superconducting properties of nanofilms, the general results of the present paper remain valid.

Nevertheless, there is another factor which can considerably affects the
shape resonances in experiments. Note that in our study we assume the hard wall potential profile
in the direction perpendicular to the film what refers to the case of the so-called free-standing
nanofilms. However, in experiments a thin film is never isolated but 
grows on a substrate (for Pb nanofilms usually on Si). Due to the interface effect the substrate
layers strongly affect both the electronic structure and the phononic dispersion of the nanofilm. 
The role of the substrate on the shape resonances has been recently studied
in Ref.~\cite{Romero2014}. Taking into account a finite lifetime of the quantized states and
modeling the substrate/thin-film interface by a more realistic finite step potential, the authors of
Ref.~\cite{Romero2014} showed that for the case of strong-coupling to the substrate the shape
resonances are significantly suppressed in reference to the free-standing limit. However, for
metallic superconductors (with long coherence length) the enhancement of the energy gap is so strong
that the shape resonances should be experimentally observed despite the destructive influence of the
substrate.

Finally, we would like to point our that one should be careful when using the BCS theory in the
description of the strongly-coupled superconductors such as Pb, as in some cases it can lead to
overestimation of the size effect with respect to the behavior reported in the
experiments~\cite{Bose2010}. The more appropriate description of the strongly-coupled
superconductors requires the use of the Eliashberg theory. Nevertheless, the results from recent
experiments~\cite{Zhang2010,Qin2009,Guo2004,Eom2006,Bose2010}
for Pb nanostructures are qualitatively well described by the BCS model.

\section*{Acknowledgments}
This work  was financed from the budget for Polish Science in the years 2013-2015. Project number:
IP2012 048572. M. Z. acknowledges the financial support from the Foundation for Polish Science (FNP)
within project TEAM.

\end{document}